\documentclass{article}
\usepackage{epsfig}

\date{\today}

\usepackage{amsmath}

\newcommand{\I}{{\mathcal I}}

\begin{document}

\title{Global monopoles, cosmological constant
\\
and maximal mass conjecture}
\author{{\large Yves Brihaye \footnote{yves.brihaye@umh.ac.be}}\\
\small{
Facult\'e des Sciences, Universit\'e de Mons-Hainaut,
B-7000 Mons, Belgium }\\
{ }\\
{\large Betti Hartmann \footnote{b.hartmann@iu-bremen.de}}\\
\small{
School of Engineering and Science, International University Bremen, 28725 Bremen, Germany
 }\\
{ }\\
{\large Eugen Radu\footnote{radu@thphys.nuim.ie}}\\
\small{
Department of  Mathematical Physics,
National University of Ireland Maynooth, Ireland}}

\date{\today}

\maketitle

\begin{abstract}
We consider global monopoles as well as black holes with global
monopole hair in Einstein-Goldstone model with a cosmological
constant in four spacetime dimensions. Similar to the $\Lambda=0$
case, the
 mass of these solutions defined in the standard way diverges.
We use a boundary counterterm subtraction method to compute the mass and action
of $\Lambda \neq 0$ configurations.
 The mass of the asymptotically
de Sitter solutions computed in this way turns out to take positive
values in a specific parameter range and, for a relaxed set of
asymptotic boundary conditions, yields a counterexample to the
maximal mass conjecture.

\end{abstract}

\renewcommand{\thefootnote}{\arabic{footnote}}
\section{Introduction}
Although there is now strong evidence that the universe has a
positive cosmological constant \cite{super}, there is also much
interest in studying theories of gravity with a negative
cosmological constant. This interest followed the anti-de
Sitter/conformal field theory conjecture, which proposes a
correspondence  between physical effects associated with gravitating
fields propagating in anti-de Sitter (AdS) spacetime and those of a
conformal field theory (CFT) on the boundary of AdS spacetime
\cite{Witten:1998qj, Maldacena:1997re}. The results in the
literature suggest also the existence of a de Sitter (dS) version of
this conjecture which has a number of similarities with the AdS/CFT
correspondence, although many details and interpretations remain to
be clarified (see \cite{Klemm:2004mb} for a recent review and a
large set of references on these problems).

In view of these developments, an examination of
the classical solutions of gravitating
fields in spacetimes with a cosmological constant seems appropriate.

The case of a Goldstone scalar field
represents a particularly interesting model.
Gravitating global monopoles without a cosmological constant
were first discussed in \cite{vile,harari}. These  topological defects
are predicted in unified theories and may appear in cosmological
phase transitions in the early universe.
The gravitating global monopoles
have a negative mass and a deficit angle depending on the
vacuum expectation value (vev) of the scalar Goldstone field and the gravitational
coupling. For sufficiently high enough values of the vev, the solutions
have a cosmological horizon \cite{lieb, Maison:1999pi}. 
These solutions were named (after their string counterparts
\cite{laguna}) ``supermassive monopoles''.

In \cite{li} and \cite{bbh} the basic features of global monopoles
in both asymptotically dS and AdS spacetimes have been studied.
However, the issue of mass computation of these solutions has not
been addressed in the literature, nor the question of cosmological
configurations with a black hole event horizon.

In this paper, we reinvestigate global monopoles in (A)dS space-time,
special attention
being paid to the solutions' mass and action computation.
For $\Lambda > 0$, we construct global monopoles in de Sitter space
in- and outside the cosmological horizon, which has to our knowledge not been
done so far.
In addition, we also study the corresponding  cosmological solutions with a black hole horizon,
which have up to now only been studied for $\Lambda=0$ \cite{lieb} 
(see also \cite{Cai:1997ij} for a discussion of the status of no hair conjecture 
for solutions with a cosmological horizon).

Although the solutions we discuss are still asymptotically (A)dS,
their standard gravitational mass diverges,
similar to the $\Lambda=0$ case.
However, asymptotically AdS solutions with a diverging ADM mass
have been considered recently by some
authors, mainly for a scalar field in the bulk
(see e.g. \cite{Hertog:2004dr}-\cite{Liu:2004it}).
In this case it was
possible to relax the standard asymptotic
conditions without loosing the original symmetries,
but modifying the charges in order to take into account the presence of matter
fields and to find a well-defined mass of the solutions.

In this paper we propose a similar mass computation for the case of
a scalar field with a spontaneously broken internal $O(3)$ symmetry.
This is relevant especially for the case of a positive cosmological
constant. The authors of \cite{Balasubramanian:2001nb}  conjectured
that ``any asymptotic dS spacetime with mass greater than dS has a
cosmological singularity'', which is known in the literature as the
\textit{maximal mass conjecture}. Roughly speaking, it means that
the conserved mass of any physically reasonable asymptotically dS
spacetime must be negative ($i.e.$ less than the zero value of the
pure dS$_4$ spacetime). Here we argue that this is valid for a
rather limited set of dS boundary conditions, the dS global
monopoles providing an explicit counterexample since their mass may
take positive values.

Our paper is organized as follows: in Section 2, we give the model
including the equations of motion and the boundary conditions. In Section 3 we present
our numerical results, while in Section 4,
we address the problem
of mass and action definition for global monopoles in (A)dS.
 Section 5 contains our conclusions.

\section{The Model}
We consider the following action principle
\begin{eqnarray}
\label{action}
I =\int_M d^4 x \sqrt{-g} \left( \frac{1}{16\pi G}(R- 2 \Lambda)
+L_m\right)
-\frac{1}{8\pi G}\int_{\partial M} d^{3}x\sqrt{|h|}K,
\end{eqnarray}
with the matter Lagrangian
\begin{eqnarray}
L_m=- \frac{1}{2}\partial_{\mu} \Phi^a \partial^{\mu} \Phi^a  -
\frac{ \lambda}{4}(\Phi^a\Phi^a- \eta^2)^2 ,
\end{eqnarray}
which describes a Goldstone triplet $\Phi^a$, $a=1,2,3$, minimally coupled to
Einstein gravity with a cosmological constant $\Lambda$.
$G$ is  Newton's constant,
$\lambda$ the self-coupling constant of the Goldstone field
 and $\eta$ the  vev of the Goldstone field.

 The last term in  (\ref{action}) is the Hawking-Gibbons surface term \cite{Gibbons:1976ue},
which is required in order to have a well-defined variational
principle, $K$ being the trace of the extrinsic curvature for the
boundary $\partial\mathcal{M}$ and $h$   the induced metric of the
boundary.

For the metric, the spherically symmetric Ansatz
in Schwarzschild-like coordinates reads~:
\begin{eqnarray}
ds^{2}=g_{\mu\nu}dx^{\mu}dx^{\nu}=
N^{-1}(r)dr^2+r^2 (d\theta^2+\sin^2\theta
d\varphi^2)-A^{2}(r)N(r)dt^2
\label{metric}
\ ,
\end{eqnarray}
with
\begin{eqnarray}
N=1-\frac{2m(r)}{r}-\frac{\Lambda}{3}r^2,
\end{eqnarray}
while for the Goldstone field with a unit winding number, we choose
the usual hedgehog Ansatz \cite{vile} 
\begin{eqnarray}
 {\Phi}^a =  \phi(r) {e_r}^a
\ ,
\end{eqnarray}
where $e_r$ is a unit radial vector.
\subsection{The equations}
Varying (\ref{action}) with respect to the metric, we obtain the
Einstein equations which can be combined to give two first order
differential equations for $A$ and $m$:~
\begin{equation}
\label{eq1}
m'=4 \pi G\left(Nr^2 \phi'^2+2 \phi^2+r^2 V(\phi)\right) \ ,
\end{equation}
\begin{equation}
\label{eq2}
A'=8 \pi G r A \phi'^2.
\end{equation}
Variation with respect to the matter field leads to the
Euler-Lagrange equation for the Goldstone scalar:
\begin{equation}
\label{eq3}
(A Nr^2 \phi')'=A \left(2 \phi+\frac{1}{2}\frac{\partial V(\phi)}{\partial
\phi}\right),
\end{equation}
with the potential $V(\phi)=\frac{ \lambda}{4}(\phi^2- \eta^2)^2$ and
the prime denotes the derivative with respect to $r$.
Note that the equations have the same structure as  for the
$\Lambda=0$ case \cite{vile,harari}. The cosmological
constant just appears in the relation defining the metric function $N(r)$.

\subsection{Boundary conditions}
In order to solve the system of equations (\ref{eq1})-(\ref{eq3}) we
have to impose appropriate boundary conditions for the metric and
scalar field.  We start by imposing the Goldstone scalar to approach
asymptotically the vev,
$\Phi^a \Phi^a (r\to \infty) \to \eta^2$,
which assures a vanishing scalar potential.

For the metric we consider a weakened set of asymptotic boundary
conditions as compared to the standard choice in the literature. In
the AdS case, following \cite{Hollands:2005wt}, we suppose that one
can attach a boundary, $\I \cong R \times S^{2}$ to $M$ such that
$\tilde M = M \cup \I$ is a manifold with boundary. On $\tilde M$,
there is a smooth metric $\tilde{g}_{ab}$ and a smooth function
$\Omega$ such that $g_{ab} = \Omega^{-2} \tilde{g}_{ab}$, and such
that $\Omega = 0$ and $ \tilde{n}_a \equiv \tilde{\nabla}_a \Omega
\neq 0 $ at points of $\I$ (a possible choice for the line element
(\ref{metric}) is $\Omega=1/r$). The metric $\tilde h_{ab}$ on $\I$
induced by $\tilde{g}_{ab}$ is in the conformal class of the
Einstein static universe, $ \tilde h_{ab} \, dx^a dx^b = e^{\omega}
[- dt^2 + \ell^2 d \sigma^2], $ where $d\sigma^2$ is the line
element of the unit sphere $S^2$, and $\omega$ is some smooth
function (with $\ell^2=3/|\Lambda|$) \footnote{These general metric boundary
conditions hold also for less symmetric solutions, $e.g.$ axially
symmetric global monopoles which may exist for a winding number $n
\neq 1$.}.

A similar set of boundary conditions is considered for solutions
with a positive cosmological constant. In this case, the metric
$\tilde h_{ab}$ on the boundary induced by $\tilde{g}_{ab}$ is in
the conformal class of the Euclidean Einstein static universe, $
\tilde h_{ab} \, dx^a dx^b = e^{\omega} [dt^2 + \ell^2d \sigma^2], $
$i.e.$ $\I$ is a spacelike boundary.

The field equations (\ref{eq1})-(\ref{eq3}) imply the following
behavior as $r \to \infty$ 
\begin{eqnarray}
\nonumber m(r )&=&8 \pi G \eta^2 r+M_0+\frac{144 \pi G \eta^4
\lambda} {(3 \eta^2 \lambda-2\Lambda)^2} \frac{1}{r}+O(1/r^3),
\\
\label{asp} 
A(r )&=&1-\frac{288 \pi G \eta^2}{(3 \lambda \eta^2 -2
\Lambda)^2} \frac{1}{r^4}+O(1/r^5),
\\
\nonumber \phi(r )&=&\eta+\frac{6 \eta}{2\Lambda-3\eta^2
\lambda}\frac{1}{r^2}+ \frac{18 \eta^3 (\lambda (96 \pi G
\eta^2-9)-64 \pi G \Lambda)} {(3 \lambda \eta^2 -2
\Lambda)^2(4\Lambda+3\eta^2 \lambda)} \frac{1}{r^4}+O(1/r^5).
\end{eqnarray}
Therefore the function $m(r)$ whose asymptotically value usually
gives the mass of solution, presents a linear divergence as $r \to
\infty$ (a similar linear divergence in the ADM mass was found also
in various gauged supergravities with a dilaton field possessing  a
nontrivial potential approaching a constant negative value at
infinity, see $e.g.$ \cite{Hertog:2004dr}, \cite{Radu:2004xp}).
~However, this is consistent with the asymptotic set of boundary
conditions we considered \footnote{Note that, similar to other
$\Lambda \neq 0$ solutions with a linearly divergent mass function
(see $e.g.$ \cite{Chamseddine:2004xu}), it is possible to redefine
the radial coordinate $r \to (1-8 \pi G \eta^2)^{1/2}\bar{r}$, such
that the metric at infinity is (A)dS with an angular deficit.
However, this transformation should be supplemented by a further
rescaling $t \to (1-8 \pi G \eta^2)^{-1/2}\bar{t}$, and would imply
a rescaling in the boundary metric as well.}.
\\
For particle-like solutions, $r=0$ is a regular origin, the
following set of initial conditions being satisfied:
\begin{eqnarray}
\nonumber m(r )&=&\frac{1}{3} \pi G(12 \phi_1^2+\lambda
\eta^4)r^3+O(r^5),
\\
A(r  )&=&A_0+4\pi G \phi_1^2 A_0 r^2+O(r^4),
\\
\phi(r  ) &=& \phi_1 r+\frac{\phi_1}{16}\left (-3 \eta^2 \lambda+16
\pi G(9\phi_1^2+\eta^4 \lambda) +8\lambda\right)r^3+O(r^4),
\nonumber
\end{eqnarray}
where $\phi_1$ and $A_0$ are free parameters to be determined numerically.

We will also consider solutions with an event horizon located for
some $r=r_h>0$ (``black holes inside global monopoles''). The
behaviour of the functions close to $r=r_h$ reads:
\begin{eqnarray}
\label{eh} \nonumber m(r)&=& m(r_h)+m'(r_h)(r-r_h)+O(r-r_h)^2,
\\
A(r )&=&A(r_h)+A'(r_h) (r-r_h)+O(r-r_h)^2,
\\
\nonumber \phi(r )&=& \phi(r_h)+\phi'(r_h)(r-r_h)+O(r-r_h)^2,
\end{eqnarray}
with
\begin{eqnarray}
\label{eh1}
m_h&=&\frac{r_h}{2}\left(1-\frac{\Lambda}{3}r_h^2\right),~~
m'(r_h)=4 \pi G \left(2 \phi(r_h)^2+r_h^2V(\phi(r_h))\right),
\\
A'(r_h)&=&8 \left. \pi G r_h A(r_h) \phi'(r_h)^2,~~
\phi'(r_h)=\frac{1}{N'(r_h)r_h^2}\left(2\phi(r_h)+\frac{1}{2}
\frac{\partial V(\phi)}{\partial \phi}\right|_{r_h}\right) \ ,
\end{eqnarray}
where $\phi(r_h)$ and $A(r_h)$ are free parameters
and $N'(r_h)=1/r_h-\Lambda r_h/3$.

The dS solutions we consider in this paper possess a cosmological
horizon located at $r=r_c>0$, where an expansion similar to
(\ref{eh}) is valid (replacing $r_h$ by $r_c$).

By going to the Euclidean section (or by computing the surface gravity)
one finds the Hawking temperature of the solutions with an horizon at
$r=\tilde{r}$ (where $\tilde{r}=r_h$ or
$\tilde{r}=r_c$) to be:
\begin{eqnarray}
\label{TH}
T=\frac{1}{\beta}=\frac{A(\tilde{r}) |N'(\tilde{r})|}{4 \pi} \ .
\end{eqnarray}

\section{Numerical results}
To perform numerical computations and order-of-magnitude estimations,
it is useful to have a new set of dimensionless variables.
Therefore we introduce the following dimensionless coordinate $x$, field $H$
and coupling constant $\alpha$:
\begin{eqnarray}
\label{resc}
     x = \eta r , \quad    H(x)=\frac{\phi(r)}{\eta}, \quad
     \alpha^2 = 4 \pi G \eta^2,
\end{eqnarray}
and replace $ \Lambda/\eta^2 \to \Lambda$.
We also introduce a new function $\mu$ in order to subtract the linear part in (\ref{asp}):
\begin{equation}
\mu(x)=\eta\left(m(r)-8\pi G \eta^2 r\right) \ .
\end{equation}
As we will argue in Section 4,
the  mass of the solution is determined by the parameter $M_0$ which enters the
asymptotics of $m(r)$;
its dimensionless value is $M_0/\alpha^2$.

\subsection{Global monopoles in (A)dS}
We start by discussing the globally regular solutions of the system of equations
(\ref{eq1})-(\ref{eq3}). First, we will recall the features of the
$\Lambda=0$ solutions, studied in \cite{vile,harari,lieb}.
For $0\leq \alpha \leq \sqrt{1/2}$ regular solutions without horizons exist.
After a suitable rescaling of the
 coordinates, one finds
that these have a solid deficit angle $\varphi_{deficit}=2\alpha^2$
(compare (\ref{asp})).
For  $ \sqrt{1/2} \leq \alpha \leq \sqrt{3/2}$ the solutions
develop a horizon at some $\alpha$-dependent value $x=x_{sm}(\alpha)$.
These solutions are the so-called ``supermassive'' monopoles studied in
\cite{lieb}.
For $1 \leq  \alpha \leq \sqrt{3/2}$ the metric function $\mu(x)$ and
thus also $N(x)$
develop oscillations in the region outside the horizon.
For $\alpha > \sqrt{3/2}$ no static solutions exist at all.

Considering asymptotically (A)dS solutions, we observe that the
$\Lambda=0$ configurations get progressively deformed by a nonzero
cosmological constant. Here we consider the cases $\alpha = 0.5$ and
$\alpha = 1.0$ only, but we believe that these represent the generic
features of the solutions.

\begin{figure}
\centering
\epsfysize=20cm
\mbox{\epsffile{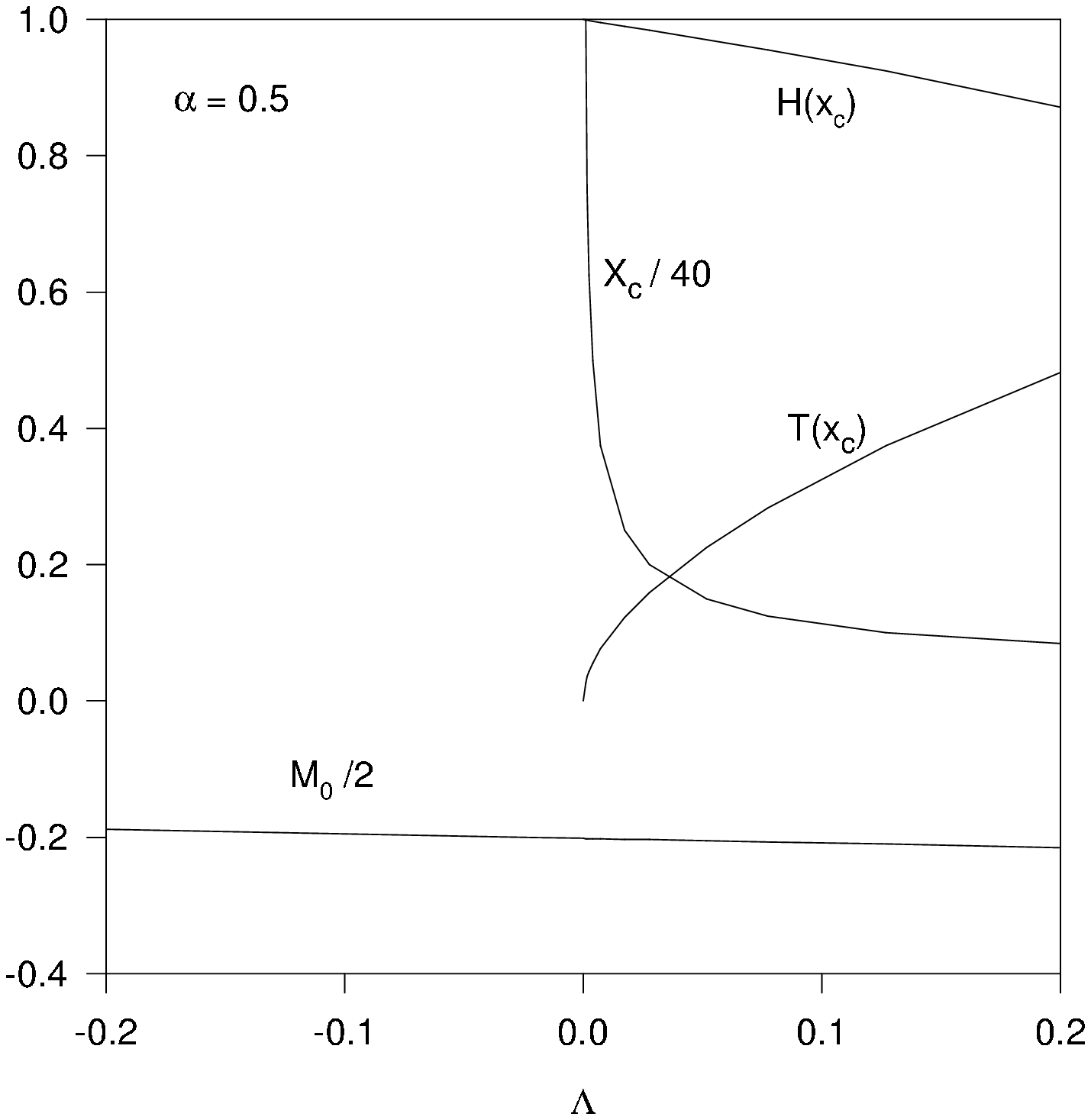}}
\caption{\label{fig1} The mass parameter $M_0$, the value of the cosmological horizon
radius $x_c$,
the temperature at the cosmological horizon $T(x_c)$ as well as the value of the
scalar field function at the cosmological horizon $H(x_c)$ are shown as  functions
of the cosmological constant $\Lambda$ for the global monopoles with $\alpha=0.5$.}
\end{figure}

\begin{figure}
\centering
\epsfysize=20cm
\mbox{\epsffile{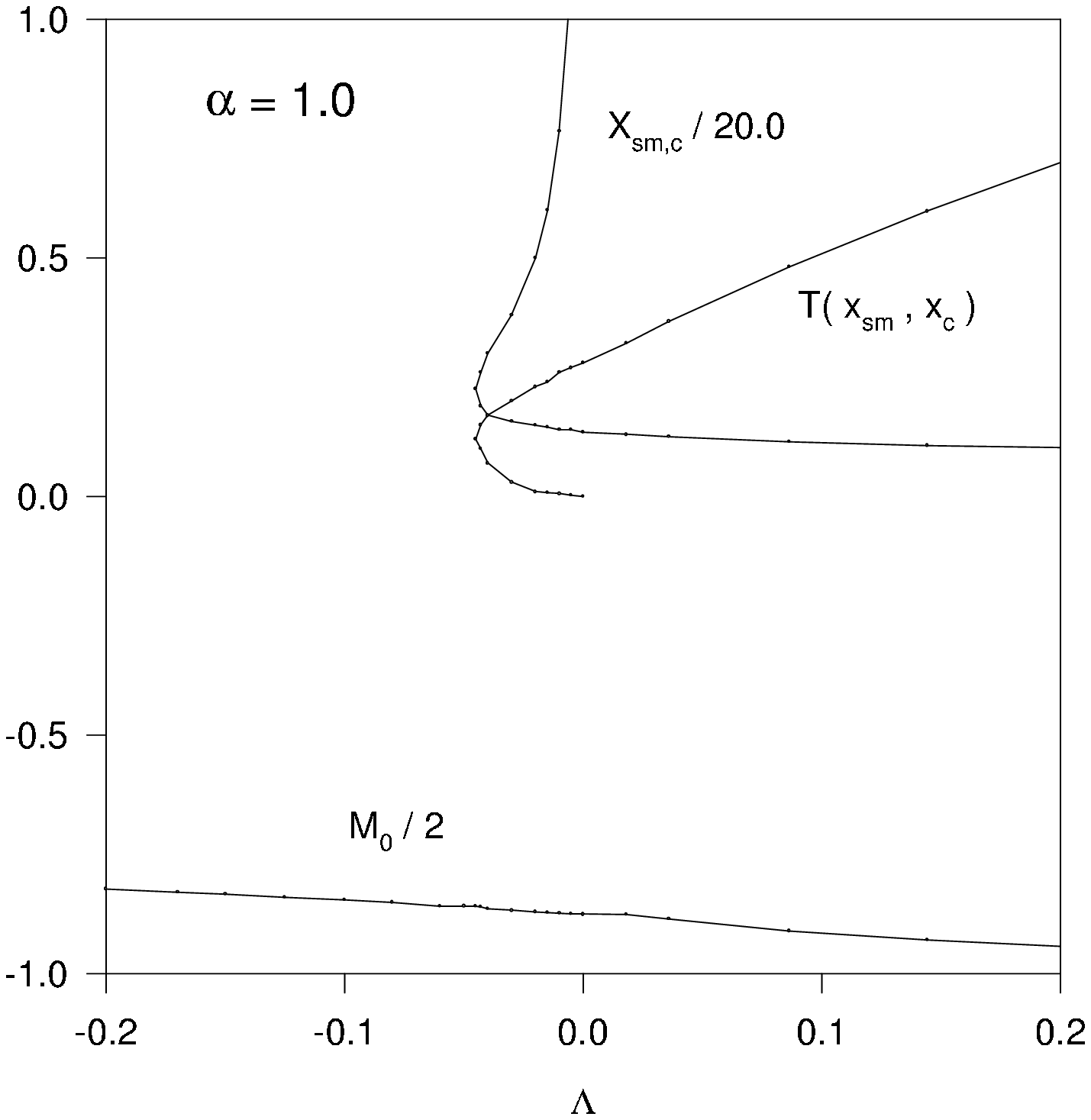}}
\caption{\label{fig2}  The mass parameter $M_0$, the value of the cosmological
horizon radius $x_c$, respectively
of the supermassive monopole horizon $x_{sm}$
as well as the temperature at the respective horizons $T(x_c,x_{sm})$
   are shown as functions of
$\Lambda$ for  global monopole solutions with $\alpha=1.0$.  }
\end{figure}

The results for these two values of $\alpha$ are
given  in Figs. \ref{fig1} and \ref{fig2}.
For $\alpha = 0.5$ and $\Lambda <0$
the global monopole exists for arbitrary values
of the negative cosmological constant.
 The parameter $M_0$ characterizing
the mass of the solution is negative. In Fig.\ref{fig1} the data
is given for $|\Lambda| \leq 0.2$, but in \cite{bbh} it is demonstrated
that $M_0$  stays negative for any $\Lambda < 0$ and approaches zero
in the  limit  $\Lambda \rightarrow -\infty$.
This feature seems to occur irrespectively of the value of $\alpha$.

For $\alpha=0.5$ and $\Lambda > 0$ the solution develops
a cosmological horizon at some
$\Lambda$-dependent value of the radial coordinate $x=x_c$.
With the boundary conditions discussed above, the procedure is to
integrate separately between the origin and
cosmological horizon and from the cosmological horizon
to infinity, matching the solutions at the cosmological horizon.
In Fig.\ref{fig1},
we plot $M_0$, the value of the cosmological horizon $x_c$,
 the Hawking temperature at the horizon $T(x_c)$ and the value of the
scalar field function $H$ at the horizon $H(x_c)$ as functions of $\Lambda$.
Again, the mass parameter
$M_0$ turns out to be always negative and decreases with increasing $\Lambda$.

The situation for $\alpha = 1.0$ is more subtle. In this case the
$\Lambda = 0$ solution possesses a ``supermassive monopole'' horizon
at  $x=x_{sm}$. Decreasing $\Lambda$ slightly from zero, we observe
that a second zero of the function $N(r)$ develops at some $\tilde
x_{sm} \gg x_{sm}$. Continuing to decrease  $\Lambda$ our numerical
analysis reveals that the values $\tilde x_{sm}$ , $x_{sm}$ approach
each other and coincide at a critical value of $\Lambda$ (for
$\alpha=1.0$ we find for $\Lambda \approx -0.045$, $x_{sm}=\tilde
x_{sm} \approx 4.5$). Then for even lower value of $\Lambda$  the
solution has no horizons. Again the mass parameter $M_0$ is
negative. For $\Lambda >0$ the solution has a single, cosmological
horizon at $x=x_c$. This zero of the metric function $N(x)$ is in
fact the continuation of the one associated with the supermassive
$\Lambda=0$ solution. The parameter $M_0$, the different horizons'
values and the temperature at these horizons are given in
Fig.\ref{fig2}.

\begin{figure}
\centering
\epsfysize=20cm
\mbox{\epsffile{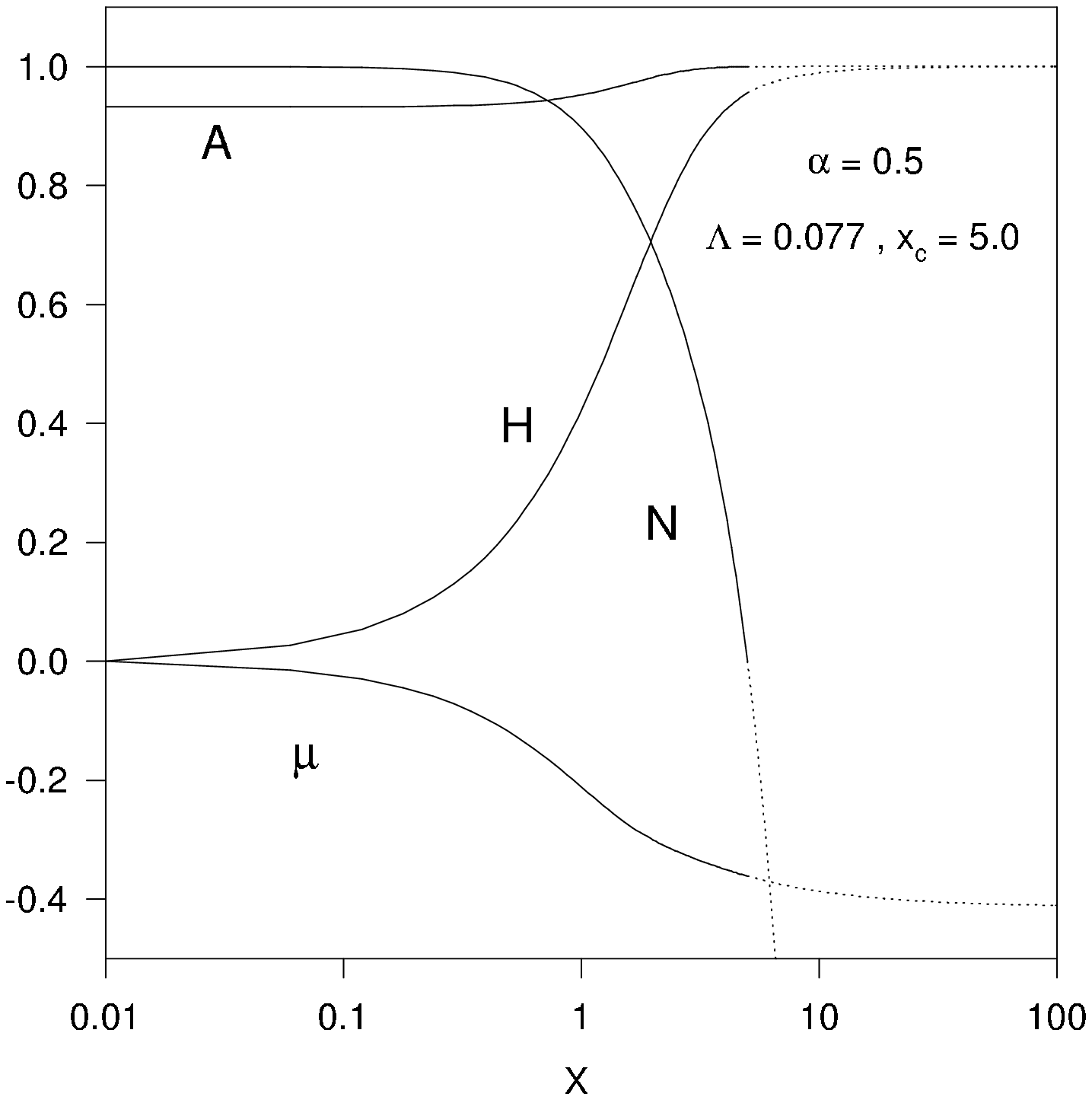}}
\caption{\label{fig3} The profiles of the metric function $A$, $N$ and $\mu$
as well as of the scalar field
function $H$ are shown for a typical global monopole in de
Sitter space with $\alpha=0.5$ and $\Lambda=0.077$.
The cosmological horizon is located at $x_c=5.0$.}
\end{figure}

\begin{figure}
\centering
\epsfysize=20cm
\mbox{\epsffile{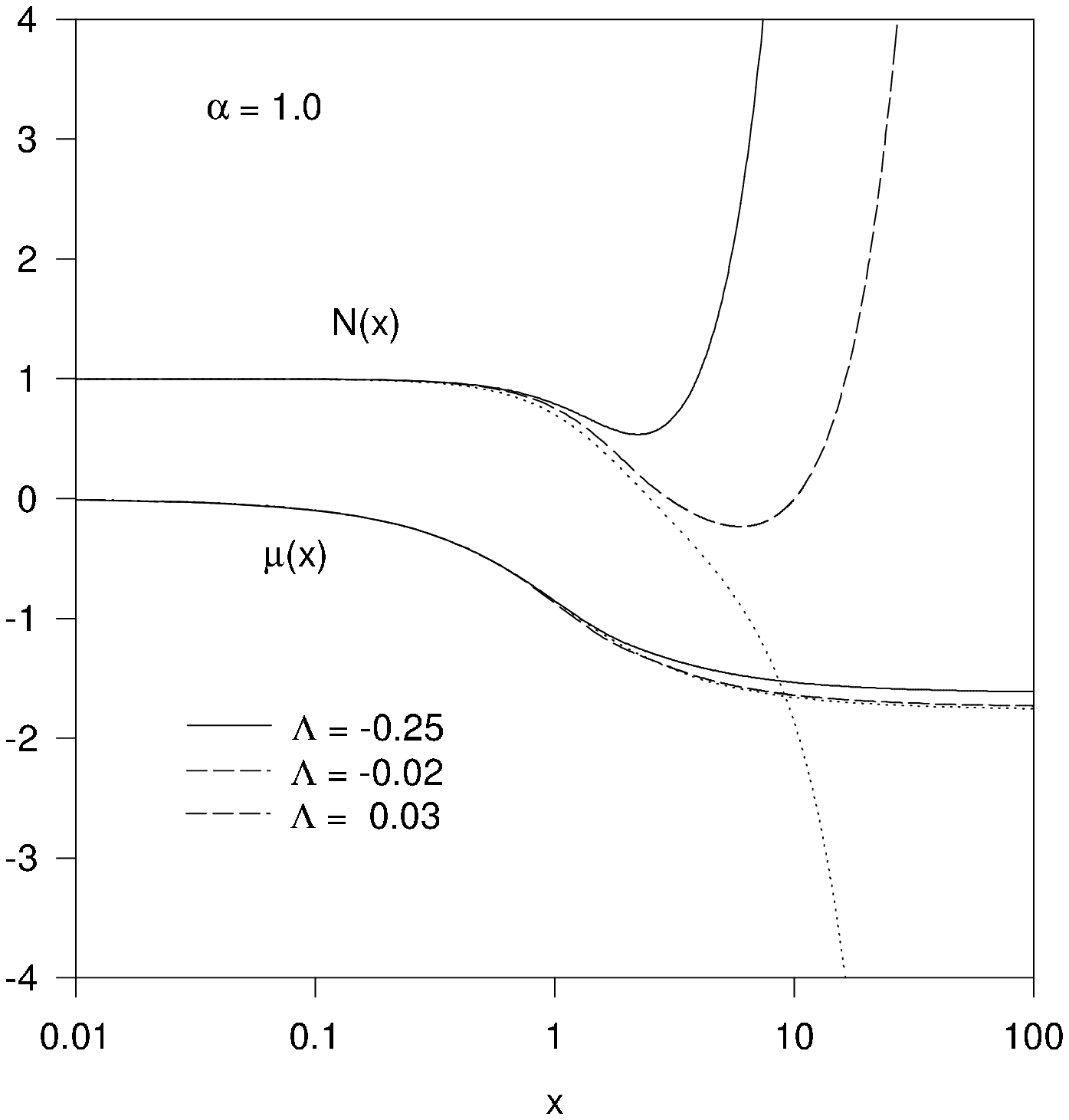}}
\caption{\label{fig4}
The profiles of the metric function $N(x)$ and
$\mu(x)$ are shown for $\alpha=1.0$
and  $\Lambda=-0.25$, $-0.02$ (AdS solutions) and
$\Lambda=0.03$ (dS solutions). }
\end{figure}

The profiles of the metric and matter functions for a typical global
monopole solution in de Sitter space are given in Fig.\ref{fig3}.
The profiles of the functions $N(x)$ and $\mu(x)$
for three characteristic values of $\Lambda$ are given
in Fig.\ref{fig4}.

We have further noticed that the oscillations of the function $N(x)$
(which are very small for $\Lambda=0$ and $\alpha=1.0$) are amplified
for increasing $\Lambda$. Therefore it becomes difficult to
accurately determine the parameter $M_0$ for $\Lambda > 0.1$. However, our numerical
results give strong evidence that it is always negative.

\subsection{Black holes inside global monopoles}
We study also black hole solutions inside global monopoles. For this, we have
to replace the boundary conditions for the global monopoles by boundary conditions
at a regular horizon located at $x=x_h$.
We find that the corresponding black
hole solutions share many features with their regular counterparts
discussed above. Since there are now three continuous parameter (namely $x_h$, $\Lambda$, $\alpha$)
to vary,
we restrict  our numerical investigations to the case $x_h=0.5$ and
to $\alpha=0.5$ and $\alpha=1.0$.

\begin{figure}
\centering
\epsfysize=20cm
\mbox{\epsffile{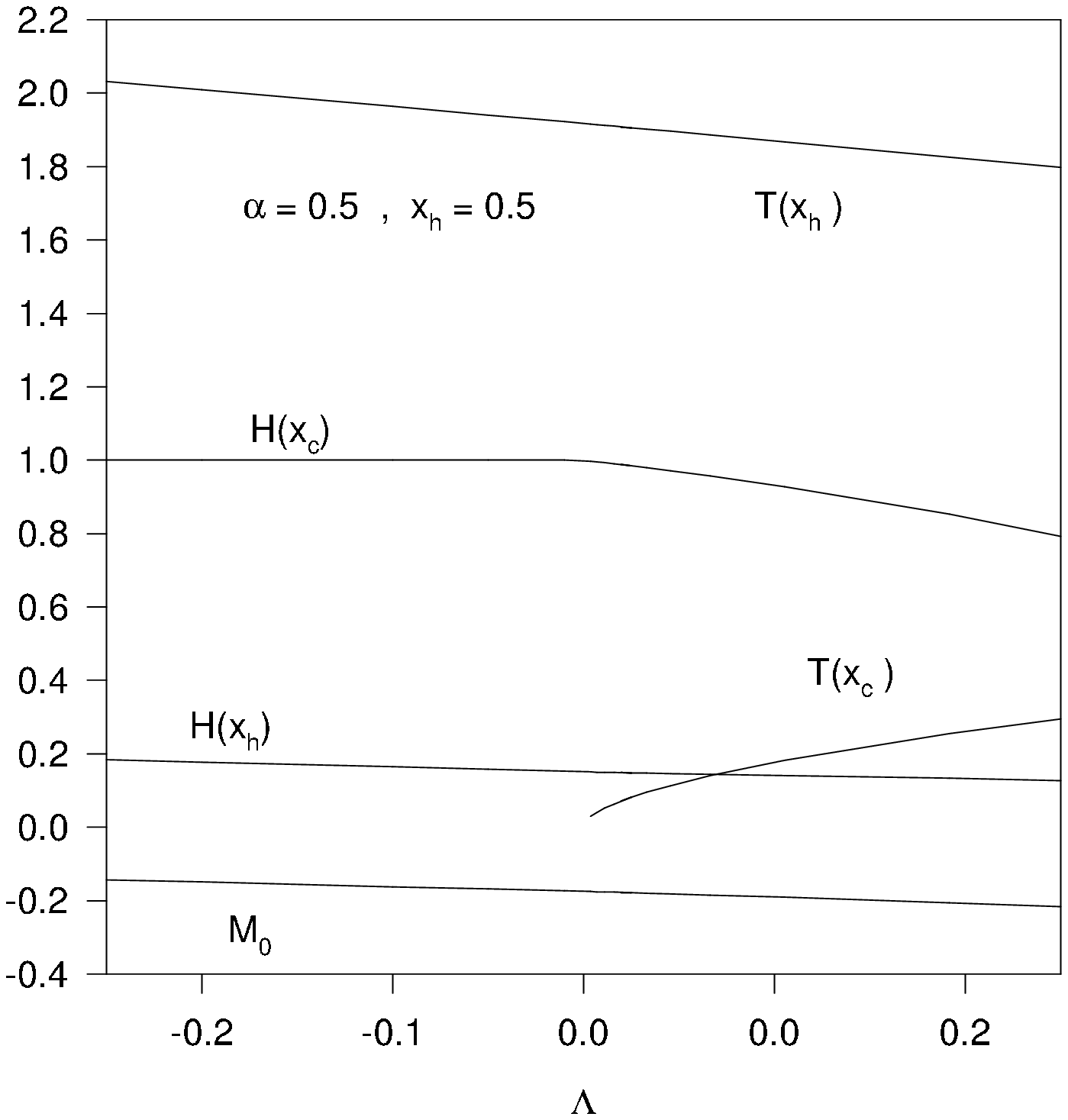} }
\caption{\label{fig5}
The values of the scalar field function $H$
and the temperature $T$ at the cosmological horizon $x_c$,  $H(x_c)$ and $T(x_c)$
as well as at the black hole horizon $x_h$, $H(x_h)$ and $T(x_h)$ are shown as functions
of $\Lambda$ for
black holes inside global monopoles with $\alpha=0.5$ and $x_h=0.5$.
The mass parameter $M_0$ is also shown. }
\end{figure}

In Fig. \ref{fig5}, we give our results for $\alpha=0.5$ and $-0.3
\leq \Lambda \leq 0.3$.  There is shown also the mass parameter
$M_0$ which in this case is again negative and decreases
monotonically with increasing cosmological constant (this was
checked for larger values of $\Lambda$ than the set shown in the
figure). In asymptotically dS spacetime two horizons occur (an event
horizon at $x=x_h$ and a cosmological horizon at $x=x_c$). The
Hawking temperatures at the two horizons $T(x_c)$ and $T(x_h)$ are
also given in Fig.\ref{fig5}.

For increasing $\Lambda$ the two horizons approach each other (the cosmological
horizon moves to smaller $x$) and for $\Lambda \sim 0.4$ we notice that
the function $\mu(r)$ becomes oscillating. As far as our numerical
simulation confirms, the mean values of the oscillations is negative
but it becomes impossible to compute the mass of the solution by the
method underlined above.

\begin{figure}
\centering
\epsfysize=20cm
\mbox{\epsffile{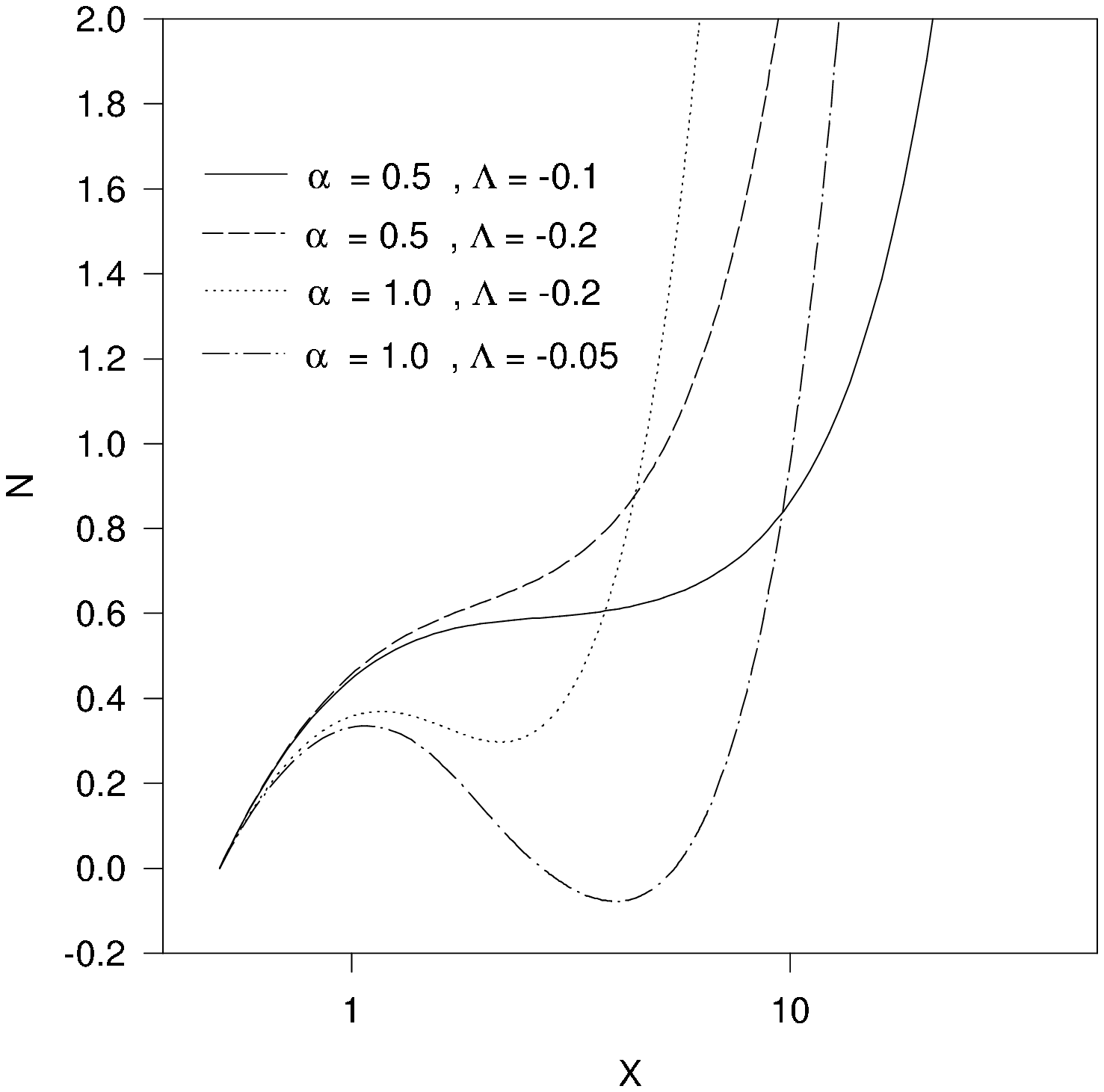}}
\caption{\label{fig7} The profiles of the metric function $N(x)$ are shown for different choices
of $(\alpha,\Lambda)$ in AdS space. Note that the solutions with $\alpha=1.0$ and
$\Lambda=-0.05$ possesses $3$ horizons.  }
\end{figure}

We find also that for small values of $\alpha$ (e.g. $\alpha=0.5$)
the evolution of the function $N(x)$ with varying $\Lambda$ shows no
occurrence of a pronounced local minimum. This is shown in Fig.
\ref{fig7}. This figure however reveals that this changes for large
enough values of $\alpha$. For $\alpha = 1$, indeed, the function
$N$ develops a local minimum for $\Lambda \geq -0.2$. While
$\Lambda$ increases the minimal value of $N$ becomes smaller and
smaller, approaching zero for some $x_m$ and then taking even
negative values. Thus, for $-0.05 \leq \Lambda \leq 0$ the function
$N$ has three zeros (see Fig.\ref{fig7}).
 We have found that the
largest of these zeros tends to infinity in the $\Lambda \rightarrow
0$ such that the solution has two zeros for $\Lambda\geq 0$.

\begin{figure}
\centering
\epsfysize=20cm
\mbox{\epsffile{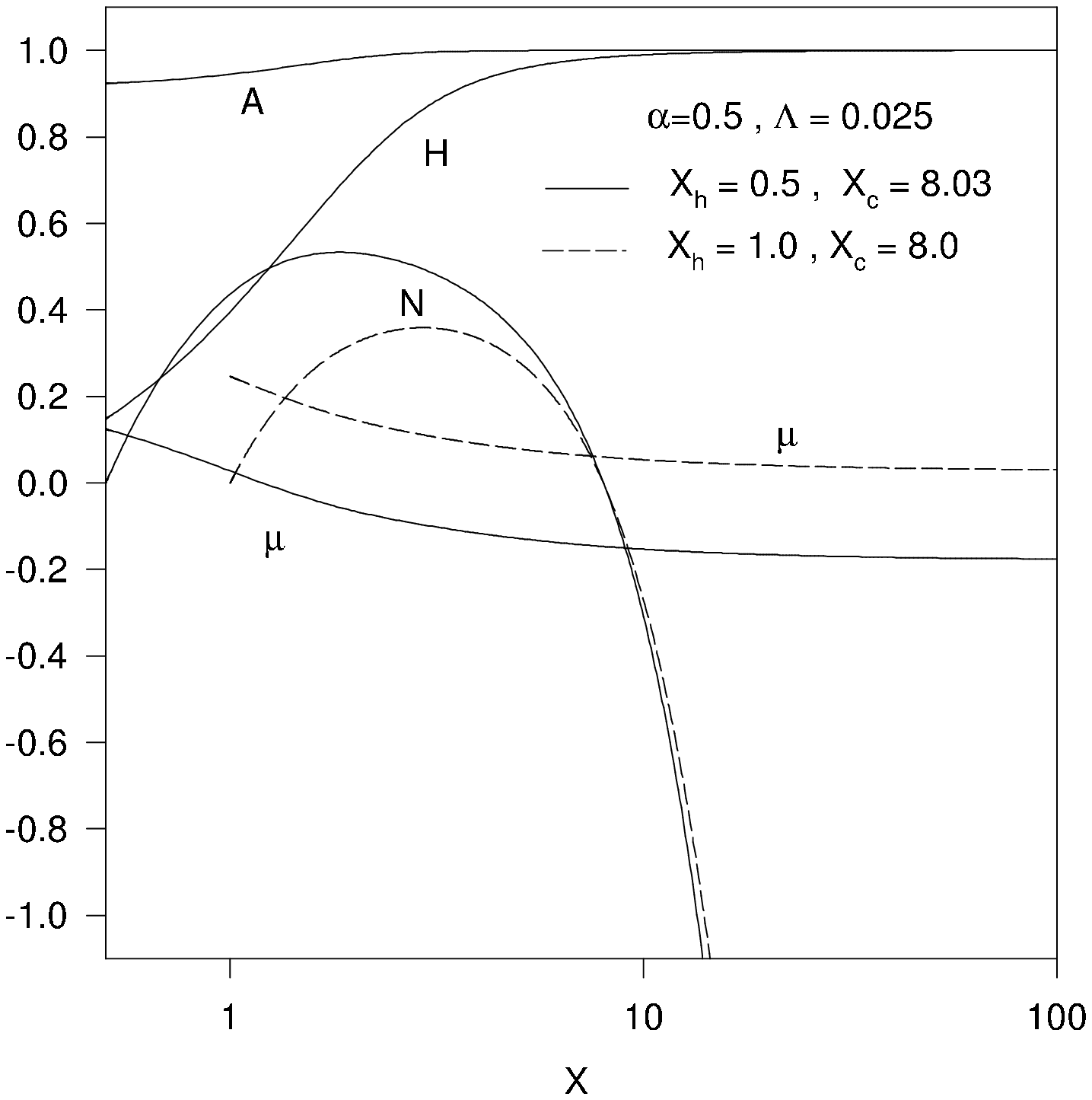}}
\caption{\label{fig6} The profiles of the metric functions $A$, $N$ and $\mu$
as well as of the scalar field
function $H$ are shown for two typical black holes inside global monopoles in
de Sitter space with $\alpha=0.5$, $\Lambda=0.025$.
Note that the profiles of the functions $A$ and $H$ for $x_h=1$ are almost identically with
those of the  $x_h=0.5$ solution.}
\end{figure}

Different from globally regular solutions, the mass parameter $M_0$
of the cosmological black hole configurations may take positive
values as well, for both signs of $\Lambda$. The occurrence of black
hole solutions with $M_0>0$ of the gravitating Goldstone model has
been already noticed in \cite{Nucamendi:2000af} for $\Lambda=0$. In
Fig. \ref{fig6} we show the profiles of two typical dS solution with
$\Lambda = 0.025$, $\alpha=0.5$ and two different values of $x_h$.
Clearly, the solution with $x_h=1.0$ has $M_0 > 0$ ($\mu$ approaches
a positive value for $x\to\infty$). We find that configurations with
mass parameter $M_0\geq 0$ exist also in the $\Lambda<0$ case, for
large enough values of the event horizon radius.

\section{Global monopole mass and action}
\subsection{AdS solutions }
An important problem of AdS space concerns the definition
of action and conserved charges of asymptotically AdS solutions.
As concerning the mass, the generalization of Komar's formula in this case is not
straightforward and
requires the further subtraction of a reference background configuration in order
to render a finite result.
This problem
was addressed for the first time in the 1980s,
with different approaches (see for instance \cite{Hollands:2005wt}
for a recent review).

In the Brown-York approach \cite{Brown:1992br}, one defines a
quasilocal stress-tensor through the variation of
the gravitational action
\begin{eqnarray}
\label{s0}
T_{ab}= \frac{2}{\sqrt{-h}} \frac{\delta I}{ \delta h^{ab}}.
\end{eqnarray}
The conserved charge associated with the
time translation symmetry of the boundary metric
is the mass of spacetime.
However, $T_{ab}$  diverges as the boundary is pushed to infinity,
and hence a background
subtraction is again necessary.
In \cite{Balasubramanian:1999re} this procedure has been improved by
regulating the boundary stress tensor
through the introduction of an appropriate boundary counterterm.
This method does not require
the introduction of a somewhat artificial reference background and it has become
the standard approach when applied to AdS/CFT, as the boundary
counterterms have a natural interpretation in terms of
conventional field theory counterterms that show up in the dual CFT.

As found in \cite{Balasubramanian:1999re},
the following counterterms are sufficient to cancel
divergences in four dimensions,
for vacuum solutions with a negative cosmological constant $\Lambda=-3/\ell^2$
(in this Section we do not take the rescaling (\ref{resc}))
\begin{eqnarray}
\label{ct}
I_{\rm ct}^{0}=-\frac{1}{8 \pi G} \int_{\partial {\mathcal M}}d^{3}x\sqrt{-h}\Biggl[
\frac{2}{\ell}+\frac{\ell}{2}\cal{R}
\Bigg]\ ,
\end{eqnarray}
where  ${\cal R}$ is the Ricci scalar
for the boundary metric $h$.

Using this counterterm, one can
construct a divergence-free boundary stress tensor $T_{\mu \nu}$
from the total action
$I{=}I_{\rm bulk}{+}I_{\rm surf}{+}I_{\rm ct}^0$ by defining
\begin{eqnarray}
\label{s1}
T_{ab}^{(0)}
=\frac{1}{8\pi G }(K_{ab}-Kh_{ab}-\frac{2}{\ell}h_{ab}+\ell E_{ab}),
\end{eqnarray}
where $E_{ab}$ is the Einstein tensor of the  boundary metric,
$K_{ab}=-1/2 (\nabla_a n_b+\nabla_b n_a)$ is the extrinsic curvature,
$n^a$ being an outward pointing normal vector to the boundary.

If $\xi^{a}$ is a Killing vector generating an isometry of the boundary geometry,
there should be an associated conserved charge.
We suppose that the boundary geometry is foliated
by spacelike surfaces $\Sigma$ with metric
$\sigma_{ij}$
\begin{eqnarray}
\label{b-AdS}
h_{ab}dx^{a} dx^{b}=-N_{\Sigma}^2dt^2
+\sigma_{ij}(dx^i+N_{\sigma}^i dt) (dx^j+N_{\sigma}^j dt).
\end{eqnarray}
The conserved charge associated with
time translation $\partial /\partial t$ symmetry of the boundary metric
is the mass of spacetime
\begin{eqnarray}
\label{mass}
\mathbf{M}=\int_{ \Sigma}d^{2}x\sqrt{\sigma}N_{\Sigma}\epsilon,
\end{eqnarray}
where $\epsilon=u^{a}u^{b}T_{a b}$ is the proper energy density
and $u^{a}$ is a timelike unit normal to $\Sigma$.

The presence of the additional matter fields in the bulk action
brings the potential danger of having divergent contributions coming
from both the gravitational and matter action
\cite{Taylor-Robinson:2000xw}. Various examples of asymptotically
AdS solutions whose action and mass cannot be regularized by
employing only the counterterm (\ref{ct}) have been presented in the
literature. 
This is also the case of the asymptotically 
AdS solutions of the Goldstone model, whose mass and action
are not regularized by
the counterterm (\ref{ct}).
To evaluate the Euclidean bulk action, we use
the Killing identity
$\nabla^a\nabla_b K_a=R_{bc}K^c,$
for the Killing field $K^a=\delta^a_t$, together with the Einstein equation
\begin{eqnarray}
\label{Rtt}
\frac{1}{16\pi G}(R- 2 \Lambda)
+L_m=\frac{1}{8 \pi G}R_t^t,
\end{eqnarray}
such that it is possible to write the bulk action
contribution as a difference of two 
surface integrals at infinity and on the event horizon, or at $r=0$ for configurations
with a regular origin.
Restricting to this last class of solutions,
one find from (\ref{asp})  $I_{bulk}=4 \pi \beta(r^3/(8\pi G \ell^2)+M/(8\pi G)+O(1/r))$,
with $\beta$ the periodicity of the Euclidean time.
(The black hole solutions have a finite supplementary event horizon
 contribution).
The Gibbons-Hawking boundary term yields
$I_{GH}=4 \pi \beta(-3r^3/(8\pi G \ell^2)-(1/(4\pi G)-4 \eta^2)r+3M/(8\pi G)+O(1/r))$.
One can see that
the geometric boundary counterterm (\ref{ct}),
$I_{ct}^0=4 \pi \beta( r^3/(4\pi G \ell^2)-(1/(4\pi G)-2 \eta^2)r-M/(4\pi G)+O(1/r))$,
fails to regularize the divergencies associated with the bulk scalar field, 
the solutions' action diverging as $I\sim 8 \pi \beta \eta^2 r$.
Also, the component $T_t^{t(0)}$ of the boundary stress tensor
decays too slow, $T_t^{t(0)}=-2 \ell \eta^2
/r^2 -M \ell/(4 \pi G r^3)+O(1/r^4)$, which
implies a divergent mass, as computed according to (\ref{mass}).

However, in such cases, it is still  possible to obtain a finite
mass and action by allowing $I_{ct}$ to depend not only on the
boundary metric $h _{ab}$, but also on the matter fields. This means
that the quasilocal stress-energy tensor (\ref{s1}) also acquires a
contribution coming from the matter fields.
There are two main prescriptions for calculating these supplementary 
boundary counterterms. The first of these involves the asymptotic
expansion of metric tensor and 
bulk matter fields near the boundary of spacetime. 
The matter counterterms are covariant expressions
of fields living at the boundary which 
remove all divergencies of the on-shell action and give finite conserved charges.
A systematic development of this method for bulk gravity coupled to scalar fields
was given first in \cite{deHaro:2000xn}
(see also \cite{Bianchi:2001kw}, \cite{Skenderis:2002wp}).

A sufficient general matter counterterm action for the four dimensional solutions
we consider here is given by
$I_{ct}^{(\phi) }=
 \int_{\partial M}d^{3}x\sqrt{-h }
(Y(\Phi)+C(\Phi){\cal R})$,
where $Y(\Phi)$, $C(\Phi)$ are polynomial functions of the Goldstone scalar.
We also suppose that $\eta$ does not enter this generic expression ($\eta$ will appear 
only when  considering the asymptotic expression (\ref{asp}) of the scalar field).
As implied by (\ref{asp}), the large $r$ asymptotics of the matter counterterm action
is $I_{ct}^{(\phi) }\sim (Y(\eta)r^3/\ell+2C(\eta)r/\ell)$. 
One can see that $Y(\Phi)=0$ necessarily, while the choice $C(\Phi)=\ell \Phi^a \Phi^a$
removes the linear divergence in the total action.

The second method is based on the Hamilton-Jacobi formalism and
was first applied in the AdS/CFT context in \cite{deBoer:1999xf}.
In this approach, one takes into account the holographic
principle of flows in the radial
direction and defines conjugate momenta to the bulk field variables
and the Hamiltonian with respect to the
AdS radial coordinate $r$.
Diffeomorphism invariance of the theory constrains
the Hamiltonian to vanish.
To obtain the Hamilton-Jacobi equation one must rewrite the Hamiltonian
constraint in terms of functional derivatives of the on-shell action.
The conjugate momenta takes a supplementary contribution 
due to the counterterm action.
The counterterms are determined
by solving the Hamilton-Jacobi equation order by order in the metric expansion
(see \cite{Martelli:2002sp} for a review of this method).

The ref. \cite{Batrachenko:2004fd} 
considered asymptotically AdS solutions for a general bosonic
action consisting in gravity
coupled to a set of scalar and vector fields,
in $d$-spacetime dimensions.
The general form of the corresponding boundary counterterms was derived 
by using the Hamilton-Jacobi formalism.
The action principle in \cite{Batrachenko:2004fd} 
is general enough to contain (\ref{action}) as a particular case. 
Thus one can use the results derived there for the equations
satisfied by $Y(\Phi)$ and $C(\Phi)$.
As expected, one finds
that the choice $Y(\Phi)=0$ and $C(\Phi)=\ell \Phi^a \Phi^a$
solves the first order expansion in powers of $\Phi^a$ of the general equations in 
Section 3 of 
\cite{Batrachenko:2004fd}\footnote{Note that the boundary counterterms equations derived in 
\cite{Batrachenko:2004fd}   
include also the geometric terms (\ref{ct}).
Also, in the concrete examples discussed there, the scalar field was supposed to vanish
asymptotically as $\Phi^a\sim1/r^{d-3}$.}.
 
Thus we find
that by adding a  counterterm of the form
\begin{eqnarray}
\label{Ict}
I_{ct}^{(\phi) }=
 \int_{\partial M}d^{3}x\sqrt{-h }
~\ell   \phi^2 {\cal R}
\end{eqnarray}
to the expression (\ref{ct}), the linear divergence associated with a divergent asymptotic
value of $m(r)$ disappears.
This yields a supplementary contribution to (\ref{s1}),
$T_{ab}^{(\phi)}=-2 \ell  \phi^2 E_{ab}$.
The nonvanishing components of the resulting boundary stress-tensor
$T_{ab}=T_{ab}^{(0)}+T_{ab}^{(\phi)}$ are
\begin{eqnarray}
\label{BD4}
\nonumber
T_{\theta}^{\theta} =T_{\varphi}^{ \varphi}=
\Big(\ \frac{M_0}{8 \pi G}\Big)\frac{\ell}{r^3}
+O\left(\frac{1}{r^4} \right),
~~
T_{t}^t=\Big( -\frac{M_0}{4 \pi G}\Big)\frac{\ell}{r^3}+O\left(\frac{1}{r^4}\right).
\end{eqnarray}
We remark that, to leading order, this stress tensor is traceless as
expected from the AdS/CFT correspondence, since even dimensional
bulk theories are dual to odd dimensional CFTs which have a
vanishing trace anomaly. Employing the AdS/CFT correspondence, this
result can be interpreted (after a suitable rescaling) as the
expectation value of the stress tensor in the boundary CFT
\cite{Myers:1999ps}. The mass of solutions, as computed from
(\ref{mass}) is
$\bf{M}$=$M_0$.

As stated above, the sum of the counterterms (\ref{ct})  and   (\ref{Ict}) regularizes the
infrared divergencies, such that
the contribution  from the asymptotic region to the total action is found to be
$\beta M_0/G$ (where $\beta$ is given by (\ref{TH})
for black hole solutions and takes arbitrary values
for regular configurations).
For black hole solutions, there is also an event horizon contribution from the bulk term,
$I_h=\frac{\beta}{4 G}  (r^2 g^{tt}g^{rr}g_{tt,r})|_{r_h}$.
Thus, the standard relation
\begin{eqnarray}
\label{entropy}
S=\beta \mathbf{M}-I
\end{eqnarray}
gives an entropy of the black holes with global monopole hair which
is one quarter of the event horizon area.

\subsection{The mass of dS solutions }
The computation of  mass
in an asymptotically dS
spacetime is
 a  more difficult task.  This is due to the absence of the spatial
 infinity and the globally timelike
Killing vector in  this case.

In \cite{Balasubramanian:2001nb}, a novel prescription
was proposed, this  obstacle being
avoided by computing the quasilocal
tensor of Brown and York (augmented by the AdS/CFT inspired counterterms),
on the Euclidean surfaces at $\mathcal{I}^{\pm}$.
The conserved charge associated with the Killing vector $\partial/\partial t$
- now spacelike outside the cosmological horizon-
was interpreted as the conserved mass.
This allows also a discussion of the thermodynamics
of the asymptotically dS solutions outside the event horizon,
the boundary counterterms regularizing the (tree-level)
gravitation action \cite{Ghezelbash:2001vs}.
The efficiency of this approach has been demonstrated in a broad range of examples.

In this approach, the initial action (\ref{action}) is supplemented
again by a boundary counterterm action $I_{ct}$, depending only on
geometric invariants of the boundary metric. In four dimensions, the
counterterm expression is (here $\Lambda=3/\ell^2$)
\cite{Balasubramanian:2001nb,Ghezelbash:2001vs}
\begin{eqnarray}
\label{actionct}
I_{ct}^0 =-\frac{1}{8\pi G}\int_{\mathcal{\partial M}^{\pm }} d^{3}x\sqrt{h }
\Biggl[ \frac{2}{\ell }%
-\frac{\ell   }{2 }\mathcal{R}
\Bigg],
\end{eqnarray}%
the corresponding boundary stress tensor being
\begin{eqnarray}
\label{s2}
T_{\mu \nu}^{(0)}
=\frac{1}{8\pi G }(K_{\mu \nu}-Kh_{\mu \nu}-\frac{2}{\ell}h_{\mu \nu}-\ell E_{\mu \nu}).
\end{eqnarray}
(Note that these expressions can be obtained from the AdS counterparts (\ref{ct}), (\ref{s1}) 
\cite{Ghezelbash:2001vs}). 
However, we have found that the function $m(r)$ always diverges as $r\to \infty$ for solutions
with a positive cosmological constant.
As a result, the counterterm (\ref{actionct}) fails to regularize
the action and mass of Goldstone-dS configurations, 
the structure of the divergences being similar to
the AdS case.

The issue of mass and action renormalization for such cases is 
considerably less explored for dS asymptotics.
However, 
we find that by supplementing the total action with
 the dS version of the matter counterterm (\ref{Ict}) 
\begin{eqnarray}
\label{Ictds}
I_{ct}^{(\phi) }=
 -\int_{\mathcal{\partial M}^{\pm }} d^{3}x\sqrt{h } 
~\ell~\Phi^2 {\cal R},
\end{eqnarray}%
yields a finite action for Goldstone-dS configurations
(evaluated at timelike infinity
outside the cosmological horizon).
This gives also a supplementary contribution to the total boundary stress tensor
$T_{ab}^{(\phi)}=2 \ell  \phi^2 E_{ab}$.

As a result, the mass of asymptotically dS configurations is
$\bf{M}$=$-M_0$, which in the absence of matter fields agrees with
the standard value for the Schwarzschild-dS solution (the overall
sign-flip in the mass arises from the relative signature change in
the boundary metric as compared to AdS case). However, we have found
numerically that the dS global monopoles have $M_0 < 0$ for a  
range of parameters, without 
the occurance of a cosmological singularity. This implies a
positive value of $\bf{M}$, contradicting obviously the  maximal
mass conjecture proposed in \cite{Balasubramanian:2001nb},
which therefore should be supplemented by specifying the  set of
boundary conditions satisfied by the metric functions at
infinity. 

\section{Conclusions}
 In this paper we have performed a general study of the properties
of the gravitating Goldstone-model solutions with a cosmological constant,
 in four spacetime dimensions.
 Both globally regular solutions and configurations with a black hole
 event horizon have been considered.

  The previous results on
globally regular solutions in dS space have been extended by constructing
the solutions both inside and outside the cosmological horizon. For large enough values
of $\alpha$, these solutions have two horizons, one relating to the horizon
of the ``supermassive monopole''  and one to the cosmological horizon.

Studying black holes inside global monopoles, we have noticed that
black holes solutions with up to three zeros of the metric function
are possible in AdS space for large enough gravitational coupling
and $\Lambda$ close to zero.
The mass parameter of the black hole solutions may take both positive and negative values.

The asymptotic behaviour of the metric in the presence of a
Goldstone scalar field is different from that in pure gravity.
Similar to the $\Lambda=0$ case, the mass-energy density of these
configurations decreases only like $1/r^2$, such that the total mass
energy defined in the usual way is diverging  linearly at large
distances. 
The mass of the asymptotically dS solutions computed according to a
counterterm subtraction procedure turns out to take positive values,
which violates the maximal mass conjecture put forward in
\cite{Balasubramanian:2001nb}.

One may argue that this violation is not a surprise, since 
 the maximal mass conjecture has been proposed in the context of 
pure dS gravity.
Its validity has been has
been tested for Schwarzschild-dS black holes \cite{Ghezelbash:2001vs}
and topological dS solutions \cite{Cai:2001jd}.
However, it has also been verified for
several different solutions with matter fields, 
including dilatonic deformation
of the dS black holes \cite{Cai:2001jd} and 
Reissner-Nordstr\"om-dS black holes \cite{Astefanesei:2003gw}.
A violation of this
conjecture has been noticed in \cite{Clarkson:2003wa} (see also
\cite{Mann:2004mi}). However, this case requires the presence of a
NUT charge and thus a different form of the boundary metric.
A review of this subject is presented in \cite{Clarkson:2004yp}.

Our results indicate that a more general matter content 
may allow, however, for configurations whose mass
is greater than that of dS spacetime.
Here we should emphasize again the central role played in our
example by the choice of asymptotic set of boundary conditions
satisfied by the metric functions.
(This choice was imposed by the particular matter model we have considered.) 
The cases discussed so far in
this context assumed  the more restrictive set $g_{tt} \sim
-1+\Lambda r^2/3+O(1/r)$, $ g_{rr} \sim -3/(\Lambda
r^2)-9/(\Lambda^2 r^4)+O(1/r^5)$, or the equivalent expression when
replacing the two-sphere with a surface with the same amount of
symmetry
 (the nut-charged case requires a separate
discussion). It would be interesting to consider the $\Lambda>0$
analogous of various asymptotically AdS systems with a divergent ADM
mass discussed in the literature and to look for further violations
of the maximal mass conjecture.

\newpage
{\bf Acknowledgments}\\
ER would like to thank  Cristian Stelea for  valuable remarks on a
draft of this paper.
 YB gratefully acknowledges the Belgian FNRS for
financial support. The work of ER is carried out in the framework of
Enterprise--Ireland Basic Science Research Project SC/2003/390 of
Enterprise-Ireland.

\newpage

\newpage

\newpage

\newpage

\newpage

\newpage

\newpage

\end{document}